\documentclass{article}

\usepackage{arxiv}

\usepackage[utf8]{inputenc} 
\usepackage[T1]{fontenc}    
\usepackage{hyperref}       
\usepackage{url}            
\usepackage{booktabs}       
\usepackage{amsfonts}       
\usepackage{amsmath}
\usepackage{nicefrac}       
\usepackage{microtype}      
\usepackage{lipsum}
\usepackage{graphicx}
\graphicspath{ {./images/} }

\title{A Simplified and Numerically Stable Approach to the BG/NBD Churn Prediction model}

\author{
 Dylan Zammit \\
  Gaming Innovation Group \\
  \texttt{dylan.zammit@gig.com} \\
   \And
 Christopher Zerafa \\
  Gaming Innovation Group \\
  \texttt{christopher.zerafaa@gig.com}
}

\begin{document}
\maketitle
\begin{abstract}
This study extends the BG/NBD churn probability model, addressing its limitations in industries where customer behaviour is often influenced by seasonal events and possibly high purchase counts. We propose a modified definition of churn, considering a customer to have churned if they make no purchases within $M$ days. Our contribution is twofold: First, we simplify the general equation for the specific case of zero purchases within $M$ days. Second, we derive an alternative expression using numerical techniques to mitigate numerical overflow or underflow issues. This approach provides a more practical and robust method for predicting customer churn in industries with irregular purchase patterns.
\end{abstract}


\section{Literature Review}
Churn prediction is a critical area of research in various domains, including customer relationship management, marketing, and finance. Accurate customer churn prediction allows businesses to address customer attrition and improve retention strategies proactively. Numerous models have been developed to address this challenge, ranging from traditional statistical methods to more advanced machine-learning techniques.

Traditional statistical methods for churn prediction often rely on survival analysis techniques, such as the Pareto/NBD model. The Pareto/NBD model, introduced by \cite{schmittlein}, is a widely used probabilistic model for customer-base analysis. It assumes that customers make purchases at a rate governed by a Poisson process, and their unobserved attrition rate follows a gamma distribution. The model allows for the estimation of customer lifetime value (CLV) and the prediction of future purchasing behaviour.

Bayesian statistical methods offer an alternative approach to churn prediction by incorporating prior knowledge and updating beliefs based on observed data. The BG/NBD model, proposed by \cite{bgnbd_hardie}, is a Bayesian counterpart to the Pareto/NBD model. It provides a more tractable framework for estimating customer churn probabilities and has been shown to yield results similar to the Pareto/NBD model while being easier to implement. Further enhancements were done by \cite{bgnbd_improved} by modelling the count of customer purchases as COM-Poisson distributed instead of Poisson distributed, giving greater flexibility. \cite{bgnbd_complaints} extend the model by incorporating customer complaints as part of the equation, positing that complaints, although infrequent, affect the probability of churn of customers significantly. In this paper, we will not be dealing with either of these two extensions, and we will only be simplifying a generalised expression in \cite{bgnbd_hardie_notes}.

A key difference between the Pareto/NBD model and the BG/NBD model is in the way a customer is defined as churned. In the Pareto/NBD model, dropout can occur at any unobservable point in time, exponentially distributed since the first activity, whereas the BG/NBD assumes that dropout occurs immediately after a purchase occurs. At either of these points, the customer becomes inactive and never makes a purchase again.

While these models have been widely applied, they face limitations in industries with highly seasonal or event-driven purchase patterns, such as the iGaming sector. In these contexts, these definitions of churn become impractical, as customers may naturally have long periods of inactivity between major sporting events or seasons. 

To address these limitations, a more practical definition of churn is required. One such approach, as mentioned in the introduction, defines churn as the absence of purchases within a specified time frame of some duration. This definition aligns better with industries where long periods of inactivity are common but do not necessarily indicate permanent customer loss. Moreover, from a business perspective, it makes sense to decide on a hard cut-off by the end of which we can categorically decide if a customer has churned or not. In other words, using our definition, a customer who is considered to have churned, might make another purchase at some point in the future, and churn again, and so on.

\cite{borg} addresses this approach in more detail, giving particular importance to new customers, which tend to have a higher churn rate than long-term customers, and hence being more important to model. In their case, a window of 12 days was used to decide whether a customer would churn or not, and the problem is treated as a binary classification problem. \cite{mrgreen}'s work, done in collaboration with Mr Green, take a similar approach, also noting that early customers are the most essential customers to take into consideration. In the latter, a 30-day window was used.

In the context of the BG/NBD model, Equation 34 in Section 5.3 of \cite{bgnbd_hardie_notes} gives a more general expression of the probability that a customer will make $y$ purchases in a period of duration $t$. This cumbersome expression involves the exponentiation of a term by the number of unique purchasing days performed throughout the lifetime of a customer. In high-activity industries, such as that of the iGaming industry, this value might be in the hundreds for players with a long history. Exponentiating by such a number results in overflow issues, leading to intractable solutions. 

To address these challenges, our research not only focuses on the simplification of this expression for the special case where $y=0$, but also employs numerical techniques to further mitigate the possibility of overflow and underflow. 

We have earlier mentioned the works of \cite{borg} and \cite{mrgreen}, applying machine learning techniques to model churn as a binary classification problem. \cite{bgnbd_application_steam} provides an application of the BG/NBD model to the online gaming industry, using Steam user activity data, \cite{bgnbd_application_mobile_gaming} applies the same model to mobile games, and \cite{bgnbd_application_banking} even applies this statistical model to the banking industry. However, to the best of our knowledge, we were not able to identify any literature related to the iGaming industry making reference to the BG/NBD model that was put forward by \cite{bgnbd_hardie}. 

An application of BG/NBD using iGaming data and a comparison of the different methods discussed in the literature would be an interesting area of research that would complement this paper well. This would provide a new modelling perspective of churn prediction in the industry, overcoming problems such as seasonal and event-driven activity inherent in the iGaming industry. This statistical model also provides a more explainable implementation, which is often required by businesses to enable informed decision-making.

\section{Derivation}
To get the probability that a customer makes no purchases within a period of duration $t$, we would need to set $y=0$ in Equation (34) of \cite{bgnbd_hardie_notes} and set the value of $t$ to the future point in time at which point $M$ days of no activity occurs. For instance, if we consider a churned customer to be someone who does not make a purchase within 30 days, and the last day of deposit $t_x$ occurred 20 days ago, then to get the probability of this customer churning, we would need to set $t=T + 10 = t_x + 30$ in the below expression, where $T$ is the present day. More generally we need to set $t$ to
\begin{equation}
    \tilde{t}:=\max{(0, M-t_x+T)}.
\end{equation}
From the above definition, taking the max implies that any customer who hasn't deposited in at least $M$ days, is guaranteed a churn probability of 1. Plugging in $y=0$ in Equation (34) immediately drops the $C$ term, giving us our definition of churn
\begin{equation}\label{eq:prob_orig}
    Pr(Y(t) = 0 | x, T, t_x, r, a, b, \alpha) = \frac{A + B}{L(r, a, b, \alpha|x, t_x, T)},
\end{equation}
where $r, a, b, \alpha$ are learned parameters,
\begin{equation}
    A = \frac{B(a+1, b+x-1)\Gamma(r+x)\alpha^r}{B(a, b)\Gamma(r)(\alpha+t_x)^{r+x}},
\end{equation}
\begin{equation}
    B = \frac{B(a, b+x)\Gamma(r+x)\alpha^r}{B(a, b)\Gamma(r)(\alpha+T+t)^{r+x}}
\end{equation}
and the likelihood function
\begin{equation}
    L(r, a, b, \alpha|x, t_x, T) = \frac{\Gamma(r+x)\alpha^r}{B(a, b)\Gamma(r)}\left(\frac{B(a, b+x)}{(\alpha+T)^{r+x}}+\frac{B(a+1, b+x-1)}{(\alpha+t_x)^{r+x}}\right),
\end{equation}
the latter being obtained from Equation 11 of the same notes. $B(\cdot, \cdot)$ and $\Gamma(\cdot)$ are the Beta and Gamma functions respectively.
Factoring out common terms in $A+B$ we get
\begin{equation}
    A+B=\frac{\Gamma(r+x)\alpha^r}{B(a, b)\Gamma(r)}\left(\frac{B(a+1, b+x-1)}{(\alpha+t_x)^{r+x}}+\frac{B(a, b+x)}{(\alpha+T+t)^{r+x}}\right)
\end{equation}
Plugging the values back in Equation \ref{eq:prob_orig} and cancelling terms, we get
\begin{multline}\label{eq:prob_simplified}
    Pr(Y(t) = 0 | x, T, t_x, r, a, b, \alpha) =\\ \frac{B(a+1, b+x-1)(\alpha+t_x)^{-(r+x)}+B(a, b+x)(\alpha+T+t)^{-(r+x)}}{B(a+1, b+x-1)(\alpha+t_x)^{-(r+x)} + B(a, b+x)(\alpha+T)^{-(r+x)}}.
\end{multline}
\section{Handling Underflow Issues}
Let us simplify Equation \ref{eq:prob_simplified} by writing it as
\begin{equation}\label{eq:prob_simple}
    P_t = \frac{B_1E^\phi + B_2F_t^\phi}{B_1E^\phi + B_2G^\phi}.
\end{equation}
The issue with computing this is due to exponentiation of the denominators by a large negative value of $\phi:=-(r+x)$, recalling that $x$ is the unique number of customer purchasing days.

Each one of the four terms are of the same form, so let us choose the first term in the numerator one without loss of generality. By taking the logarithm and exponentiating, we can rewrite this as
$$B_1E^\phi = \exp({\log({B_1E^\phi})})=\exp{\left[\log(B_1) + \phi\log{(E)}\right]}.$$
By applying this transformation, $\phi$ is no longer part of the exponent. However, we still have to exponentiate with respect to $K_E:=\log(B_1) + \phi\log{(E)}$, which remains a large negative integer. By repeating this process for the other terms we can rewrite Equation \ref{eq:prob_simple} as
\begin{equation}
    P_t = \frac{e^{K_E}+e^{K_{F_t}}}{e^{K_E}+e^{K_G}}.
\end{equation}
Now, let $K:=\max(K_E, K_{F_t})$ and $K':=\max(K_E, K_G)$. Thus we can rewrite $P_t$ once more as 
\begin{equation}
    P_t = e^{K-K'}\left(\frac{e^{K_E-K}+e^{K_{F_t}-K}}{e^{K_E-K'}+e^{K_G-K'}}\right).
\end{equation}
Through this transformation, all exponents of this equation are now much smaller, making the computation of this expression much more numerically stable, helping to avoid numerical underflow.

In summary, by writing down all the terms in the order in which they should be computed, we get
\begin{align}
\begin{split}
    E &:= \alpha+t_x \\
    F_t &:= \alpha+T+t \\
    G &:= \alpha+T \\ \\
    K_E &:= \log(B(a+1, b+x-1)) - (r+x)\log{(E)} \\
    K_{F_t} &:= \log(B(a, b+x)) - (r+x)\log{(F_t)} \\
    K_G &:= \log(B(a, b+x)) - (r+x)\log{(G)} \\ \\
    K &:= \max(K_E, K_{F_t})    \\
    K' &:=\max(K_E, K_G) \\ \\
    P_t &:= e^{K-K'}\left(\frac{e^{K_E-K}+e^{K_{F_t}-K}}{e^{K_E-K'}+e^{K_G-K'}}\right).
\end{split}
\end{align}
\section{Conclusion}
In this paper, we give an alternate definition of what it means for a customer to churn, providing a simplified expression of the special case of Equation 34 in \cite{bgnbd_hardie_notes}. Moreover, we offer a simple transformation involving exponentials and logarithms, providing numerical stability and tractability. This expression was implemented in the open-source GitHub repository \cite{pymc_nmarketing}.

Feature improvements include exploring applications of this model to real-world data, particularly that of the iGaming industry due to the sparsity of research of this industry using the BG/NBD model.

\bibliographystyle{unsrt}  


\end{document}